\documentclass[11pt,fleqn]{article}   

\usepackage{latexsym,bm}
\usepackage{amsmath,amsfonts,amssymb}
\usepackage{cite}

\catcode`\@=11
\@addtoreset{equation}{section}
\def\theequation{\arabic{section}.\arabic{equation}}
\def\appendix{\renewcommand{\thesection}{\Alph{section}}\setcounter{section}{0}
              \renewcommand{\theequation}
            {\mbox{\Alph{section}.\arabic{equation}}}\setcounter{equation}{0}}

\def\maketitle{\thispagestyle{empty}\setcounter{page}0\newpage
                \renewcommand{\thefootnote}{\arabic{footnote}}
                  \setcounter{footnote}0}
\renewcommand{\thanks}[1]{\renewcommand{\thefootnote}{\fnsymbol{footnote}}
               \footnote{#1}\renewcommand{\thefootnote}{\arabic{footnote}}}
\newcommand{\preprint}[1]{\hfill{\sl preprint - #1}\par\bigskip\par\rm}
\renewcommand{\title}[1]{\begin{center}\Large\bf #1\end{center}\rm\par\bigskip}
\renewcommand{\author}[1]{\begin{center}\Large #1\end{center}}
\newcommand{\address}[1]{\begin{center}\large #1\end{center}}

\def\dinfn{\smallskip Dipartimento di Fisica, Universit\`a di Trento\\ 
                           and Istituto Nazionale di Fisica Nucleare,\\
                                   Gruppo Collegato di Trento, Italia}

\def\Idinfn{\address{\dinfn}}

\newcommand{\email}[1]{e-mail: \sl #1@science.unitn.it\rm}

\newcommand{\femail}[1]{\thanks{\email{#1}}}

\def\babs{\hrule\par\begin{description}\item{Abstract: }\it} 
\def\eabs{\par\end{description}\hrule\par\medskip\rm}
\renewcommand{\date}[1]{\par\bigskip\par\sl\hfill #1\par\medskip\par\rm}
 
\newcommand{\s}[1]{\section{#1}}

\def\hs{\qquad}               
\def\beq{\begin{eqnarray}}    
\def\eeq{\end{eqnarray}}      
\def\at{\left(}               
\def\aq{\left[}               
\def\ct{\right)}              
\def\cq{\right]}              
\def\R{{\hbox{{\rm I}\kern-.2em\hbox{\rm R}}}}   
\def\H{{\hbox{{\rm I}\kern-.2em\hbox{\rm H}}}}   
\def\N{{\hbox{{\rm I}\kern-.2em\hbox{\rm N}}}}   
\def\C{{\ \hbox{{\rm I}\kern-.6em\hbox{\bf C}}}} 
\def\Z{{\hbox{{\rm Z}\kern-.4em\hbox{\rm Z}}}}   
\def\ii{\infty}                                  
\def\X{\times\,}                                  
\renewcommand{\Im}{\mathop{\rm Im}\nolimits}       
\def\al{\alpha}
\def\be{\beta}

\def\de{\delta}

\begin{document}

 \preprint{hep-th/03}
 
\title{Asymptotics of Quasi-normal Modes for Multi-horizon Black Holes }
\author{ Luciano Vanzo\femail{vanzo} and
Sergio Zerbini\femail{zerbini}}
\Idinfn
\today
\babs
The issue concerning rigorous  methods  recently developed in 
deriving the asymptotics of quasi-normal modes  is revisited  and
applied to a generic non rotating  multi-horizon black holes
solution. Some examples are illustrated and the single horizon  cases
are also considered. As a result, the asymptotics for large angular
momentum parameter is shown 
to depend on the difference between the maximal or Nariai black hole
mass and the ordinary black hole mass. The extremal limit is also
discussed and the exact evaluation of the quasi-normal frequencies
related to the Nariai space-time is presented, as a consistent check
of the general asymptotic formula.

\eabs

\s{Introduction}

Recently there has been a renewed interest in a long standing issue in 
classical relativistic  theory of gravitation: small perturbations or
 quasi-normal modes (QNMs) 
associated with static (eventually stationary) solutions of Einstein equation 
with spherical symmetry 
(for an introduction, see for example \cite{frolov,nollert,kokkotas}). 
The interest is both at fundamental level and phenomenological one, in
this last case through the generation and detection of gravitational
radiation.  

At the fundamental level, which  we will mainly interested in, it has been 
recently conjectured a connection between the real part of the QNMs frequency 
and the level spacing of the black hole area spectrum \cite{hod} (see also 
\cite{kunstatter} who considered the higher dimensional generalization along
the line of \cite{beke} and \cite{seta}). 
The idea, which can be traced back to York's
paper \cite{york:83}, was that the QNMs are to be connected with the
quantum spectrum of black hole excitations, which has to be discrete
since the entropy is finite. In an asymptotically flat context,
such conjectures have been used in the loop quantum gravity approach in order
to uniquely fix \cite{dreyer}the Immirzi parameter, which  otherwise,
within that approach remains undetermined \cite{immirzi}. 
In the presence of negative cosmological constant, an asymptotically
AdS context, a detailed analysis of the QNM spectrum for conformally
coupled scalar waves has been given for Schwarzschild-Anti-de Sitter
(SAdS) and the BTZ black holes\cite{mann1,cardlemo2001}, and later
extended to topological black holes, as well\cite{mann2}. Moreover,
there is a connection with perturbations of CFT 
quasi-equilibrium states\cite{horo:00} and their relaxation times, via
the AdS/CFT correspondence. As far as the asymptotic form of the QNMs
frequencies in the SAdS background is concerned, a rather complete
study can be found in \cite{CKL:2003}, and for Reissner-Nordtr\"{o}m
AdS black holes in \cite{bk03}.  

We recall that for 4-dimensional Schwarzschild black hole, the 
asymptotics for scalar perturbation in the limit of large
damping\footnote{A numerical investigation of the asymptotics of
  highly damped QNMs in Kerr space-time is given in \cite{bckh,bk,bcy}.}
reads  \cite{nol,and}: 
\beq
\omega_{n}\simeq 2\pi T_H
\left[i\left(n-\frac{1}{2}\right)+\frac{1}{2\pi}\ln 3\right]
+{\cal O}[n^{-1/2}]\,,
\label{as}
\eeq
where $T_H$ is the Hawking temperature of the black hole. 
The analytic derivation of this result has been presented  in 
\cite{motl,motl1,neitzke}. Indeed, the factorization of the Hawking 
temperature seems
to suggest the validity of the connection between the QNMs asymptotics and 
the physics of quantum black holes. This seems to hold true for 
a large class of space-time with a single horizon (see the recent papers
\cite{m1,pad,car,m2} and references therein).

In this paper, we would investigate with a different but mathematically 
well established method, the asymptotic for QNMs associated with a generic 
multiple horizon D-dimensional black hole, in the limit of large
angular momentum index $l$. 
Working in arbitrary dimensions ($D >3$) may be justified by the 
phenomenological interest in extra spatial dimensions 
which has recently appeared in the literature, triggered by string 
theory considerations.

The method we will make use of, is based on the so 
called ``complex scaling' or `` dilatation analytic methods'' 
(see for example \cite{briet,gerard,barreto,reed} and
original references therein). The methods, as we will see, works very well
when one is dealing with multi-horizon black holes. For this reason, we 
will consider in detail the D-dimensional Schwarzschild-de Sitter black 
hole. Several  papers have  been appeared dealing with the same issue and
making use of different approaches
\cite{moss,abdalla,lemos1,sun,zhidenko,futa,kono0}. 

The method we will make use of does not work directly for 
asymptotically AdS black holes,
but it is also appropriate  for the asymptotically flat black holes, thus
Schwarzschild and Reissner-Nordstrom black hole may be investigated.

The paper is organized as follow. In Section 2, the D-dimensional 
Schwarzschild-de Sitter  black holes is revisited and the Nariai critical mass 
is introduced. In Section 3, the master equation associated with the 
gravitational perturbation as well as the related QNMs boundary conditions are 
presented. In Section 4, the analytic dilatation approach is briefly 
summarized.  In Section 5, the asymptotics of QNMs frequencies are calculated
for a generic non rotating black hole. In Section 6 and 7, the general formula
is applied to the   D-dimensional Schwarzschild and  Schwarzschild-de Sitter  
cases. In Section 8, the exact formula of QNMs frequencies for the 
D-dimensional Nariai case is obtained and the consistency with the previous
result checked. The Conclusion and an Appendix on the extremal limits
will end the paper.

\section{D-dimensional Schwarzschild-de Sitter  Black Holes}

We recall that the D-dimensional Einstein Eqs. in vacuum with
cosmological constant are given by 
\beq
\label{eins}
R_{\mu\nu}-\frac{1}{2}R\,g_{\mu\nu}+\Lambda g_{\mu\nu}=0\,.
\eeq
A generic metric representing a non 
rotating black hole solution of the above equations, in the static 
coordinates (Schwarzschild gauge) reads
\beq
ds^{2} = -A(r)dt^2 + A^{-1}(r)dr^2+ r^2
h_{ij}dx^{i}dx^{j}\,,
\label{BHS}
\eeq
where the coordinates are labeled as $x^{\mu} = (t,r,x^{i}), (i=1,...,D)$. 
The metric  $h_{ij}$ is a function of the coordinates $x^{i}$
only, and we shall refer to this metric as the horizon
metric. In more than $5$-dimensions Einstein's equations
\eqref{eins} do not imply that the 
horizon be a constant curvature sub-manifold, but they do imply that
for positive $\Lambda$ the horizon is compact. Hence we take 
it to be a closed orientable ${D-2}$-dimensional
manifold. Black hole  solutions may be  defined by   functions $A(r)$ 
having simple and positive zeros. For metrics of the form \eqref{BHS}
these zeros determine various horizons loci.

When the cosmological constant
is absent or negative, typically one may have an interior Cauchy
horizon but only one event horizon, defined by the  
largest positive root of the lapse function $A(r)$, and the range of
the allowed $r$ is an infinite interval. Let us denote by $r_H$ the
horizon radius, such that $A(r_H)=0$ with  $A'(r_H)$ not vanishing.

When the cosmological constant is positive, 
the case we are mainly interested in, there exists the possibility of 
multiple event horizons and the range of $r$ may be a  finite interval.

Near the horizon we have the expansion
\beq
A(r)\simeq A'(r_H)(r-r_H)+..\,. 
\eeq

If one considers the Euclidean version, $t=-i \tau$, and this euclidean time 
is assumed to have a period $\beta_H$,  defining the new coordinates, 
$\rho$ and $\theta$, by means of 
\beq
r=r_H+\frac{A'(r_H)}{4}\:\rho^2\,,\qquad\qquad
\tau=\frac2{A'(r_h)}\:\theta\:,
\eeq
we can write
\beq
ds^2 \simeq \rho^2d\theta^2 + d\rho^2 
+ r_H^2 h_{ij}dx^{i}dx^{j},.
\eeq

The first two terms define the metric of a flat $2$-dimensional space
in polar coordinates. The metric will be smooth, and the space
isometric to a flat disk, if and only if the period of the angle is
$2\pi$. This leads to  
\beq
\beta_H=\frac{4\pi}{|A'(r_H)|}\,.
\label{ht1}
\eeq 
where $\beta_H $  is now interpreted as inverse of the Hawking's
temperature. One often refers to this case as to the ``on-shell black
hole''. 

The general case of a D-dimensional Schwarzschild-de Sitter black hole
is described by the following lapse function,   
\beq
A(r)=1-  \frac{c_DM}{r^{D-3}}-\frac{ r^2}{L^2}\,,
\label{ad}
\eeq
where $c_D$ depends on the D-dimensional Newton constant
$G_D$ and reads 
\beq
c_D=\frac{16 \pi G_D}{(D-2)V_H}\,,
\label{cd}
\eeq
$V_H$ the volume of the horizon, a $(D-2)$-dimensional sphere of
unit radius, and $L$ is the fundamental
length scale associated with the positive cosmological constant
according to 

\beq
\Lambda=\frac{(D-1)(D-2)}{2\,L^2}\,.
\eeq

The parameter $M$ may be identified with the mass of the
 black hole as an excitation over empty de Sitter
 space\cite{halyo:02,odin}, and satisfies the first law
\beq
\de M=\frac{\kappa_H}{8\pi G_D}\,\de A_H=-\frac{\kappa_C}{8\pi G_D}\,\de A_C
\label{laws}
\eeq
where $\kappa_C$ and $A_C$ are the surface gravity and the area of the
cosmological horizon, with radius $r_C>r_H$.
The inverse temperature of the black hole is a function of the horizon
radius 
\beq
\beta_H = 2\pi\kappa_H^{-1}=\frac{4 \pi L^{2}r_H}{L^2(D-3)-(D-1) \,r_H^2}.
\label{betator}
\eeq
The inverse temperature of the cosmological horizon is given by a
similar expression
\beq
\beta_C = 2\pi\kappa_C^{-1}=\frac{4 \pi L^{2}r_C}{(D-1) \,r_C^2-L^2(D-3)}.
\label{betatcos}
\eeq
and away from the extremal case, one always has $\be_H<\be_C$.
Since the black holes admit a temperature, they possess an entropy,
the Beckenstein-Hawking black hole entropy. From \eqref{laws} and the
expression of the temperature one derives immediately the area law for
the event horizon. Or we can write the first law of black hole
thermodynamics as 
 \beq
S_{BH}=\int \beta_H dM=\int \beta_H \frac{dM}{dr_H} dr_H\,.
\eeq
On the other hand, from $A(r_H)=0$, we have
\beq
\frac{dM}{dr_H}=\frac{1}{c_DL^2}  r_H^{D-4} \aq (D-3)L^2-(D-1)
\,r_H^{2}\cq \,.  
\eeq
As a result, we get
\beq
S_{BH}=\frac{4 \pi}{c_D} \int r_H^{D-3} dr_H=
\frac{4 \pi}{(D-2)c_D}  r_H^{D-2}=
\frac{V_H}{4 G_D}  r_H^{D-2}\,.
\eeq
The disadvantage of this method is that it is not obvious how to
handle the constant of integration. One quarter of the area is also
the entropy attributed to the cosmological horizon, but in this case
it is much less obvious how this can be true.
 Nevertheless it has been shown\cite{davies} that the validity
of the generalized second law would be seriously challenged, were it
not for the geometric entropy of the cosmological horizon. 

As an illustrative example, let us consider the 4-dimensional 
Schwarzschild-de Sitter case. Here, we have
\beq
A(r)=1-\frac{2MG}{r}-\frac{\Lambda r^2}{3}\,.
\label{a1}
\eeq
For 
\beq
0 < 1- 9M^2 G^2\Lambda < 1\,,
\label{a2}
\eeq
there exist two real simple positive roots $r_H$ and $r_C>r_H$ of 
the Eq. $A(r)=0$, and the static region is $ r_H <r < r_C\,,$.
 $r_H$ corresponds to the black hole event horizon and $r_C$ to the
cosmological horizon. The area law  associated with the inner horizon
is 
\beq
S_{BH}=\frac{4 \pi r_H^{2}}{4 G}\,.
\eeq
Furthermore, there exists a maximum allowed mass value
given by
\beq
M_N=\frac{L}{3\sqrt{3}G}\,.
\label{n4}
\eeq
which represents the largest black hole one can have in de Sitter
space. This picture carries over to the D-dimensional case. The lapse
function and the Hawking's temperature read 
\beq
T_H = \frac{L^2(D-3)-(D-1)r_H^2}{4 \pi r_H L^{2}}.
\label{tor}
\eeq
\beq
A(r) = \frac{L^2r^{D-3}-c_DML^2-r^{D-1}}{L^{2}r^{D-3}}.
\label{ator}
\eeq
The critical mass may be obtained when the event horizon radius
coincides with 
the cosmological horizon, namely when there is a double zero in the
lapse function, equivalent to the vanishing  of $T_H$. Thus, 
the critical radius $r_N=r_H=r_C$ reads
\beq\label{critr}
r_N=\at \frac{D-3}{D-1}\ct^{\frac{1}{2}}L\,.
\eeq 
and the critical mass 
\beq
M_N=\frac{2}{c_D(D-3)}\at
 \frac{D-3}{D-1}\ct^{\frac{D-1}{2}}L^{D-3}=\frac{2}{(D-3)c_DL^2}r_N^{D-1}\, 
\label{cm}
\eeq
As before, for $0<M<M_N$ there are two positive roots $r_H$ and $r_C>r_H$ of 
the Eq. $A(r)=0$, and there is a static region $ r_H <r < r_C\,,$. For
$M<0$ the solution is a naked singularity surrounded by a cosmological
horizon. For $M=M_N$ the two roots coalesce and we have a Nariai
solution, the largest black hole one can have in de Sitter space. For
$M>M_N$ there are no positive real roots, $A(r)<0$ and we have a truly
naked singularity, with no horizon of any sort protecting it.    

Let us briefly comment on the meaning of this critical mass. One can
approach the extremal limit, where $M=M_N$ and the original static
region is lost, in at least two inequivalent ways\footnote{That is,
  one obtains non diffeomorphic manifolds.}. There exists a smooth
extremal limit with positive temperature, and this is the so called
D-dimensional Nariai solution, a solution of Einstein equation with
topology $dS_2 \times S_{D-2}$ and product metric. We present a 
derivation of this result in the Appendix, along the lines of ref.
\cite{caldarelli }. And there is also a singular extremal solution with zero
temperature which is in the same topological sector as the original
black hole, and which may be the more natural ground state to
consider. Very recently, a complete survey of these extremal solutions
has been presented in \cite{lemos2}.

\section{The Master Equations}

It is well known that gravitational perturbations of a black hole are of 
special interest, since they
are related to generation and detection of gravitational waves 
\cite{frolov,kokkotas}. 
One could also consider a massless scalar or vector field evolving in a fixed
black hole background.  

Within the class of  black holes we are interested in, the horizon
manifold is an higher dimensional sphere and spherical dimensional
reduction is always  
possible. Thus, one is interested in the radial part of the perturbation 
equation. It is also well known that, instead of working with the original 
radial coordinate $r$, it is 
more convenient to introduce a new coordinate, called Regge-Wheeler or 
tortoise coordinate, denoted by $x$, defined by
\beq
x=\int \frac{dr}{A(r)}\,,\hs \frac{dx}{dr}=A(r)\,.
\label{a3}
\eeq
Since near the black hole horizon one has
\beq\label{near}
A(r) \simeq A'(r_H)(r-r_H)\,,
\eeq
it follows
\beq
x \simeq \frac{1}{A'(r_H)}\ln (r-r_H)\,, \hs r \simeq r_H+e^{A'(r_H) x}\,.
\eeq

Thus, the tortoise mapping maps the horizon to $-\infty$ and, for
$\Lambda=0$, spatial infinity to 
infinity. With $r_C$ replacing $r_H$, Eq.~\eqref{near} also holds near
the cosmological horizon,  
but since in this case $A^{'}(r_C)<0$, $x$ maps the finite 
open interval $r_H<r<r_C$ of the static region into the whole real
line, that is the horizon locus $r=r_C$ is mapped to $x=+\ii$.

The situation is different for asymptotically AdS black holes, which are  
relevant within the so called AdS/CFT correspondence. We may illustrate
the issue in the simplest case, namely the  BTZ 3-dimensional black hole.
In the non rotating case, we have (here using units in which $8G=1$)
\beq
ds^2=-A(r)dt^2+\frac{dr^2}{A(r)}+r^2d\theta^2\,,\,\,\,\,\,\,\,
A(r)=\frac{r^2}{L^2}-M\,.
\eeq
The horizon radius is $r_H=L\sqrt{M}$ and Hawking temperature 
$T_H=\frac{\sqrt{M}}{2 \pi L}$. The relation between $r$ and the tortoise
$x$ is known and reads
\beq
r(x)=r_H \coth \frac{\pi r_H x}{L^2}\,.
\eeq
Thus, the $r$ interval $(r_H, +\infty)$ is mapped to
$(-\infty,0)$. More generally, in AdS space the lapse function
diverges as $\simeq r^2$ at infinity, making the integral \eqref{a3}
convergent. We will comment on the implication of this fact later.

The radial differential equation for the gravitational perturbations 
around the Schwarzschild
original $D=4$ black hole solution are well known and were derived by Regge 
and Wheeler and Zerilli. 

The analog of Regge-Wheeler and Zerilli equations have been derived
recently for D-dimensional  Schwarzschild- de Sitter space-time and they are 
summarized by a
master equation (see for example \cite{giappa,kono1}). This equation 
 has the form
\beq
\left[-\frac{d^2}{dx^2}+V_{l}(r(x))\cq \phi_l(x)=\omega^2 \phi_l(x)\,,
\label{Z}
\eeq
where $\omega$ is the oscillation frequency of the perturbation,
decomposed as 

\beq
\phi_l(x)e^{-i\omega t}
\eeq
 and the potential $V_{l}$ depends on  
$l$, the index of spherical harmonics and on the tensorial nature of the
perturbation. The general form is given in  \cite{giappa,kono1}.
For our purposes, we may write  for vector and tensor perturbations 
respectively
\beq
V_{l,V}(r)=\frac{A(r)}{r^2}\aq l(l+D-3)+W_V(r)\cq \,,
\label{av}
\eeq
\beq
V_{l,T}(r)=\frac{A(r)}{r^2}\aq l(l+D-3)+W_T(r)\cq \,,
\label{at}
\eeq
where $W_V(r)$ and $W_T(r)$ are known functions independent on $l$.
For $D=4$, the vector contribution reduces to the axial or Regge-Wheeler 
contribution, while the tensor contribution is formally equal to the 
contribution associated with a scalar field.
Finally, the so called ``scalar'' perturbation potential is very complicated.
It is the one that reduces to the polar or Zerilli contribution in $D=4$.
Again, for our purposes, it will be sufficient to  write it in a  form
valid for very large $l$, namely
\beq
V_{l,S}(r)=\frac{A(r)}{r^2}l(l+D-3)\aq 1+O(\frac{1}{l^2})\cq\,.
\eeq

The Sturm-Liouville problem associated with the QNMs consists in the 
eigenvalue equation in the real line plus the so called ingoing and 
outgoing boundary condition at the infinity, namely

\beq
\phi(x) \rightarrow e^{-i\omega x} \,\,\,\,\,\,\,\,\mbox{as}\,\,\, x 
\rightarrow -\infty\,.
\label{in}  
\eeq

\beq
\phi(x) \rightarrow e^{+i\omega x}\,\,\,\,\,\,\,\,\mbox{as}\,\,\,
 x \rightarrow \infty\,,
\label{out}  
\eeq
Such boundary conditions render the formally symmetric operator of 
Eq.(\ref{Z}) 
not self-adjoint. Thus, one has, in general, complex frequencies $\omega^2$.
The real part of the QNMs frequencies are associated with the frequencies of 
the signal while the imaginary part is  related its decay in time. 
As a consequence,  these boundary condition render also highly non trivial
the computation of the frequencies and sophisticated numeral techniques have 
been invented to deal with this problem \cite{nollert,kokkotas}.

The situation is different for asymptotically AdS black holes and we recall 
that  the $r$ interval $(r_H, +\infty)$ is mapped in $(-\infty,0)$. 
For this, reason, one is forced to impose a specific boundary condition 
(Dirichlet, for example\cite{horo:00}, or vanishing flux
\cite{birm:02}) at $x=0$, namely at $r=\infty$. This is related to  
the well known fact that AdS is not a globally hyperbolic space-time.  

We conclude this section with the useful analogy with the associated 
scattering problem.  
It was soon recognized \cite{vish} that the QNMs problem is very similar to 
the one related to the scattering resonances in (non relativistic) 
quantum mechanic. Within this approach, a  rigorous treatment of the 
4-dimensional Schwarzschild case has been presented in \cite{ba}.  

Since the master potential is (exponentially) vanishing
at infinity of the real line, the asymptotically acceptable solution 
 for scattering off the barrier from the right, may be written in the
 form  
\beq
\phi(x) \simeq C(\omega)e^{-i\omega x}\,\,\,\,\,\,\,\,\mbox{as}\,\,\, x 
\rightarrow -\infty\,, 
\label{as1}
\eeq
and  
\beq
\phi(x) \simeq E(\omega)e^{-i\omega x}+F(\omega)e^{+i\omega x}
\,\,\,\,\,\,\,\,\mbox{as}\,\,\,\, x 
\rightarrow \infty\,\,.   
\label{as2}
\eeq
The scattering amplitude may be defined by
\beq
S(\omega)=\frac{F(\omega)}{E(\omega)}
\eeq
The ingoing boundary condition gives $C(\omega)$ finite and not vanishing 
and the outgoing boundary
condition gives $ E(\omega)=0$.  It turns out that only  complex $\omega$ 
satisfy such condition, namely the complex frequencies of the QNMs 
are associated with the poles of the analytic continuation of the 
scattering matrix (see, for example \cite{new}). 

\section{Analytic Dilatation  Approach}

We have recalled that the QNMs frequencies can be obtained by means of
an analytic 
continuation in the variable $\omega$. The analytic continuation will plays
also an important role in the following technique we are going to briefly 
introduce.

In the investigation of the scattering resonances
several approximation techniques have been developed. Fortunately, the one
we will make use of, the so called  analytic dilatation approach, can
be used for a large class of black hole  solutions.

Let us begin with the definition of QNMs which is useful for our purposes.
First, if one identify the QNMs as scattering resonances associated with the 
Master operator 
\beq
L= -\frac{d^2}{dx^2}+V_l(r(x))\,.
\eeq
The QNMS frequencies are the poles of the meromorphic continuation for
complex $\omega$ of the associated resolvent (Green function).

Let us introduce the scaled  operator 
\beq
L_\theta= -e^{-2\theta}\frac{d^2}{dx^2}+V_l(e^{\theta} x)\,,
\label{do}
\eeq
where $\theta$ is a complex number such that $ 0 < \Im \theta <
\frac{\pi}{2}$. 
As a consequence $L_\theta$ is {\it not} self-adjoint  and the 
eigenvalue problem 
\beq
L_\theta  \Psi(x)=\lambda \Psi(x)\,,
\eeq
admits solutions with complex $\lambda$, and this complex eigenvalues are,
of course, the poles of the associated resolvent   $ (L_\theta -z)^{-1}$.

Now, for a class of so called dilatation analytic potentials and for 
$ 0 < \Im \theta < \frac{\pi}{2}$ 
 it follows that 
these complex eigenvalues coincide with the resonance (QNMS) 
frequencies related to
the original $V_l(x)$. This fact holds true as soon as   the 
potential   is exponentially vanishing  at $ \pm \infty $. As a result, 
for black holes solution with two horizons, the hypothesis are satisfied and we
can look    for the  complex eigenvalue problem 
\beq
 e^{-2\theta}\aq -\frac{d^2}{dx^2}+e^{2\theta} V_l(e^{\theta} x)\cq \Psi(x)
=\omega^2 \Psi(x)\,.
\label{ceing}
\eeq
In particular, we may select $\theta=\frac{ i\pi}{4}$.  Thus
we have to solve the eigenvalue equation associated with a  non symmetric 
differential operator
\beq
\aq -\frac{d^2}{dx^2}+i V_l( \sqrt{i} x)\cq \Psi(x)
=i \omega^2 \Psi(x)\,.
\label{c1eing}
\eeq

Some remarks are in order. First, since
the eigenfunctions $\psi(x)$ are well behaved for large $|x|$ then the
original one cannot belong to the  Hilbert space $L_2(R)$ because
of out-going and in-going boundary conditions. This fact should help  the 
numerical calculations.
In general, the explicit solutions of this equation are not explicitly 
known and one has to make use of approximation methods. 
The analytic method can also applied
 to the case of black hole solution with one horizon, provide they are 
asymptotically flat. In fact, in this case, the vanishing of the master 
potential for $x$ going to $\infty$ is still sufficient for the
success of the approach.

\section{Asymptotics of QNMs Frequencies}

In general, the master potential as a function of $x$ is non negative and 
admits a local maximum, which is in correspondence with the maximum attained by
$V_l$ as a function of $r$. Furthermore, $V_l$ consists of two parts, one
depending on $l$ and the other depending on the spin $s$ of the
perturbation. 

 In the large $l$
limit, we have an universal dependence which is independent on the
spin $s$ and reads
\beq
V_l(r)\simeq \frac{A(r)}{r^2}l(l+D-3)\,.
\eeq
This term has an interesting meaning. In fact, if one investigates the 
equation of motion for a massless particle in the generic black hole 
background, it turns out that the classical potential has the form 
$\frac{A(r)}{r^2}J^2$, where $J$ is proportional to the classical 
angular momentum. From now, on we will work in the limit of large $l$.
The maximum is reached for $r_0$ such that
\beq
r_0\,A'(r_0)=2A(r_0)\,.
\eeq

In this limit, we may expand the potential around the maximum, thus 
\beq
V_l(r(x))\simeq V_l(r_0)+\frac{1}{2}A^2(r_0) \frac{d^2}{dr^2}V_l(r_0)
(x-x_0)^2+..\,. 
\eeq
Putting $y=x-x_0$, the Sturm Liouville equation for the QNMs frequencies reads
\beq
\aq -\frac{d^2}{dy^2}-\Omega_l^2  y^2\cq \Psi(y)
\simeq \at \omega^2-V_l(r_0) \ct \Psi(y)\,,
\label{ceingv}
\eeq
where we have put 
$$
\Omega^2_l=\frac{1}{2}A^2(r_0) |\frac{d^2}{dr^2}V_l(r_0)|
$$
The corresponding equation for the dilated potential can be simply obtained
by the substitution $y \rightarrow \sqrt{i} y$. As a result, we arrived at
\beq
\aq -\frac{d^2}{dy^2}+\Omega_l^2  y^2\cq \Psi(y)
=i \at \omega^2-V_l(r_0) \ct \Psi(y)\,.
\label{ceingv1}
\eeq
On the left hand side we have the harmonic oscillator operator. As a 
consequence, we have
\beq
\omega^2_{l,n}\simeq V_l(r_0)-i (2n+1)\Omega_l\,.
\eeq
For large $l$,  the above expression can be written in the final form
\beq
\omega_{l,n}\simeq \frac{\sqrt{A(r_0)}}{r_0}\aq \pm(l+\frac{D-3}{2})-
i (n+\frac{1}{2}) \sqrt{|\frac{r_0\,A''(r_0)}{2}- A(r_0)|}\cq\,.
\label{asymp}
\eeq

Some remarks are in order. The QNMs frequencies occur symmetrically with 
respect the imaginary axis and have a finite dimensional 
multiplicity  associated with (D-2)-dimensional spherical
harmonics. For example,
for $D=4$, the multiplicity is  $2l+1$. The formula derived above 
should be compared with 
an analog formula derived with the WKB methods (see \cite{will}. 
The results are very similar, but the 
derivation is  more simple and on a mathematically rigorous basis and
gives, for large $l$, the QNMs asymptotics in an explicit way. 
The asymptotic formula is also valid for asymptotically
flat black holes. 

It is possible to go beyond the harmonic or 
quadratic approximation, and to deal with anharmonic oscillator and related
perturbation series (see \cite{zasla}. For higher WKB order see, for example,
\cite{kono}.

\section{D-dimensional Schwarzschild Case}

In the following, we shall make  applications of the general asymptotic 
formula we have derived.
As first example, let us consider the pure D-dimensional Schwarzschild case.

Putting $r_S^{D-3}=c_DM$, we have 
\beq
A(r)=1-  \at\frac{r_S}{r}\ct^{D-3}.
\label{adnotq}
\eeq
The horizon is $r_H=r_S$ and the Hawking temperature 
$T_H=\frac{D-3}{4\pi r_H}$. Furthermore, the maximum is attained at
\beq
r_0=\at \frac{D-1}{2}\ct^{1/(D-3)}r_S=\at
\frac{(D-1)\,c_D\,M}{2}\ct^{1/(D-3)}\, 
\eeq
and we have
\beq
\omega_{l,n}\simeq \at \frac{D-3}{D-1} \ct^{1/2}\frac{1}{r_0}
\aq \pm \at l+\frac{D-3}{2}\ct-
i\at n+\frac{1}{2}\ct(D-3)^{\frac{1}{2}} \cq\,.
\label{asymp2}
\eeq
In terms of Hawking temperature we have (see the interesting paper
\cite{bcg03} for related formulas in the case of a radially infalling
particle) 
\beq
\omega_{l,n}\simeq 
\at \frac{2^{D-2}}{(D-1)^{\frac{D-1}{2}}} \ct^{\frac{1}{D-3}}
\frac{2 \pi}{\sqrt{D-3}}T_H
\aq \pm(l+\frac{D-3}{2})-
i (n+\frac{1}{2})(D-3)^{\frac{1}{2}} \cq\,.
\label{asymp2t}
\eeq

In the pure 4-dimensional Schwarzschild,
$A(r)=1-\frac{2GM}{r}$. Here $r_H=2GM$, $r_0=3GM$, and one recovers 
\cite{press,ferrari} 

\beq
\omega_{l,n}\simeq \frac{8 \pi T_H}{3 \sqrt{3}}\aq \pm (l+\frac{1}{2})-
i (n+\frac{1}{2}) \cq\,.
\label{asymp1}
\eeq
One can see that the asymptotics depends on the Hawking temperature. On
dimensional grounds, this is quite natural, since the only length in the game 
is the horizon radius, which depends on the mass M and on the Plank length 
via $G_D$. 

\section{D-dimensional Schwarzschild-de Sitter Case}

As a second example, we consider the Schwarzschild-de Sitter
 D-dimensional case. We have 
\beq
A(r)=1-  \at\frac{r_S}{r}\ct^{D-3}
-\frac{ r^2}{L^2}\,.
\label{adnotq1}
\eeq
Typically we have two horizons, the event and the cosmological horizon and
the associated Hawking temperatures. Furthermore, the maximum in the master 
potential is attained  again at
\beq
r_0=\at \frac{D-1}{2}\ct^{\frac{1}{D-3}}\,r_S\,.
\eeq
Recalling the critical radius
\beq
r_N=\at \frac{D-3}{D-1}\ct^{\frac{1}{2}} L\,,
\eeq
we have 
\beq
\omega_{l,n}\simeq \frac{1}{L}\at 
\frac{r_N^2}{r_0^2}-1 \ct^{1/2}
\aq \pm (l+\frac{D-3}{2})-
i (n+\frac{1}{2})(D-3)^{1/2} \cq\,.
\label{asympp}
\eeq
This expression can be rewritten in terms of the Nariai critical mass $M_N$ 
and the mass $M$ of the black hole. One has
\beq
\omega_{l,n}\simeq   \frac{1}{L} 
\at \at \frac{M}{M_N}\ct^{\frac{2}{D-3}} -1\ct^{\frac{1}{2}}
\aq \pm (l+\frac{D-3}{2})-
i (n+\frac{1}{2})(D-3)^{\frac{1}{2}} \cq\,.
\label{asymp3}
\eeq

For $D=4$, we recover \cite{barreto}

\beq
\omega_{l,n}\simeq \frac{\at 1-9\Lambda M^2 \ct^{1/2}}{3\sqrt{3}GM}
\aq \pm (l+\frac{1}{2})-
i (n+\frac{1}{2}) \cq\,.
\label{asymp4}
\eeq

Some remarks are in order. Here in the game enters several fundamental
lengths  
and one can see the dependence on $L$ related to the cosmological
constant and the ratio between the mass and the critical mass. In the
extremal limit, namely when 
$M \rightarrow M_N$, the asymptotics for large $l$ is formally vanishing. 
This has been recently observed in \cite{lemos1}, and may be related
to the fact that in this limit the temperature vanishes, and the
resulting space is a natural ground state.  

However, as already mentioned, there exists another extremal limit solution,
namely the D-dimensional Nariai solution, which can be obtained making use of 
the extremal limit discussed in the Appendix.  One can introduce  a Nariai time
$t_1=\varepsilon t$. Thus, the corresponding Nariai frequencies can be obtained
by means of $\omega^1=\frac{\omega}{\varepsilon}$ in the limit $\varepsilon 
\rightarrow 0$. Here, $\varepsilon$ is the extremal parameter. Let us 
investigate the extremal limit of the QNMs asymptotics (\ref{asymp3}).
We have (see Appendix)
\beq
M=M_N\at 1-k\varepsilon^2 \ct\,.
\eeq
As a result,
\beq
\at \at \frac{M}{M_N}\ct^{\frac{2}{D-3}} -1\ct^{\frac{1}{2}}
=\sqrt{\frac{2 k}{D-3}}\varepsilon \at 1+O(\varepsilon^2) \ct\,.
\eeq
As a consequence, the asymptotic behaviour of the QNMs Nariai frequencies is
predicted to be
\beq
\omega^1_{n,l}\simeq \frac{1}{L}\sqrt{\frac{2 k}{D-3}}
\aq \pm (l+\frac{D-3}{2})-
i (n+\frac{1}{2})(D-3)^{\frac{1}{2}} \cq\,.
\label{bu}
\eeq
We will check this result in the next Section.

\section{The D-dimensional Nariai Case}

In this Section, we will compute exactly the QNMs frequencies
associated with 
the $D$-dimensional Nariai solution. To begin with, by a simple
rescaling of coordinates, it is allowed to set $2k=D-1$ in the Nariai
metric \eqref{namet}, so we may write 
\beq
ds^2=-\at 1- \frac{(D-1)y^2}{L^2} \ct dt_1^2+
\frac{dy^2}{\at 1- \frac{(D-1) y^2}{L^2} \ct }+
r_N^2 dS^2_{D-2}\,,
\eeq
where $r_N$ is given by \eqref{critr}. The manifold topology is $dS_2
\X S_{D-2}$. Since $dS_2$ is not simply connected, this space-time has
two horizons at  
$$
y_H=\pm \frac{L}{\sqrt{D-1}}\,,
$$  and 
$$
-\frac{L}{\sqrt{D-1}} < y <\frac{L}{\sqrt{D-1}}\,. 
$$
$y$ just covers one half of the two dimensional hyperboloid. To both
horizons one may associate the same Hawking temperature 
$$
T_H=\frac{\sqrt{D-1}}{2\pi L}\,.
$$
 As already mentioned, the QNMs frequencies 
for this space-time can be analytically computed. In fact, the radial part of 
a scalar field or the tensor perturbation equation reads
\beq
\aq -\frac{d^2}{dx^2}+\frac{\lambda^2_\al}{r_N}
\at  1- (D-1)\frac{y^2(x)}{L^2}\ct \cq \Psi(x)
= \omega^2 \Psi(x)\,.
\label{ds2}
\eeq
where  $\lambda^2_\al=l(l+D-3)$ are the eigenvalues related to the 
$(D-2)$-dimensional 
spherical harmonics and the tortoise coordinate is
\beq
x=\frac{1}{2\pi\, T_H}\ln \at \frac{y_H+y}{y_H-y}\ct\,.
\eeq
As a result, $-\infty < x< \infty$ and
\beq
y=y_H\tanh( 2 \pi T_H x)\,.
\eeq
The radial master equation becomes
\beq
\aq -\frac{d^2}{dx^2}+
\frac{U_0}{\cosh^2\,\gamma \,x} \cq \Psi(x)
= \omega^2 \Psi(x)\,,
\label{ds21}
\eeq
where 
\beq
U_0=\frac{\lambda^2_\al}{\,r^2_N}=\frac{(D-1)\lambda^2_\al}{(D-3) \,L^2}\,,
\eeq
and
\beq
\gamma=\frac{\sqrt{2\,k}}{L}= 2\pi T_H \,.
\eeq
This equation is formally equivalent to a one-dimensional Schroedinger 
equation with a P\"{o}schl-Teller potential. This is an exact quantum
mechanics  
potential and the resonances frequencies associated with it are well known
(see for example, \cite{ferrari}). 

The exact QNMS frequencies turn out to be 
\beq
\omega_{n,\al}=\frac{\sqrt{D-1}}{L}\at 
\pm \sqrt{\frac{\lambda^2_\al}{D-3}-\frac{1}{4}}-i(n+\frac{1}{2})\ct\,.
\label{nf1}
\eeq
Recalling that $2\,k=D-1$, in the large $\lambda_\al$ limit, this expression 
coincides with the one 
(\ref{bu}) obtained previously.
The above expression can be also rewritten in terms of the Hawking temperature
\beq
\omega_{n,\al}=2\pi T_H \at 
\pm \sqrt{\frac{\lambda^2_\al}{D-3}-\frac{1}{4}}-i(n+\frac{1}{2})\ct\,.
\label{nf}
\eeq

\section{Conclusions}

The analytic dilatation method has been applied 
in order to derive the asymptotic for large angular momentum index of 
D-dimensional multi-horizon black holes. In general, the asymptotics  depends 
on the black hole
 physical parameters and when $\Lambda=0$, for dimensional reasons, the 
QNMs frequencies depends on the Hawking temperature and the Plank length,
namely the Newton  
constant. When there are other physical parameters, like the cosmological 
constant $\Lambda$ or the charge $Q$, the dependence is more subtle and the 
supposed connection between the real part of the QNMs frequency and the 
area  quantization of the black hole is much less transparent (see also 
\cite{kono0}).     

In the case of multiple horizon, we have derived a general asymptotic
formula for the QNMs frequencies. This formula, making use of a suitable 
extremal limit,  has permitted to derive an asymptotics 
formula for the Nariai space-time, which has been confirmed by an exact 
calculation. As a by product, we have confirmed that the so called 
P\"{o}schl-Teller approximation really is a near-horizon approximation, as
stressed  also recently by several authors
\cite{ferrari,moss,lemos1}. The other extremal solution, corresponding
to a zero temperature state, formally has vanishing QNMs
asymptotics. One may wonder what are the implications of this fact. A naive
conclusion might be that only a finite number of QNMs frequencies survive in 
this limiting case. However, if one takes care of the extremal 
limiting procedure, we have shown that there exist a QNMS  asymptotics that 
coincides with the Nariai space-time. On the other hand, it has been
conjectured  that in  de Sitter and asymptotically de Sitter
manifolds, quantum gravity effects should be described within a finite
dimensional Hilbert space, so that there could be the
possibility to deal with a system having not only a finite number of
degrees of freedom, but also a finite number of independent quantum
states \cite{banks,witten}. Thus if QNMs prove to be related with the
quantum spectrum of black holes, a finite dimensional Hilbert space is
ruled out. If they are to be thought as collective excitations of a
macroscopic collapsed object, then we see no contradictions with Banks
proposal. 

\section{Appendix}

In this Appendix, following \cite{caldarelli}, we shall study the extremal 
limit of a generic BH solution
 corresponding to a D-dimensional 
charged or neutral black hole depending on parameters as the mass $m$, charges
 $Q_i$ and the  cosmological constant $\Lambda$, generally denoted by $g_i$.  
These extremal limits have been investigated in \cite{GP,BH,Z,MS,MMS,NNN}.
Recall that in  the Schwarzschild static coordinates, we have 
\beq
ds^2=-A(r)dt^2+
\frac{1}{A(r)} dr^2+r^2 d\Sigma_{D-2}^2\,.
\label{RN}
\eeq
Here, $d\Sigma_{D-2}^2$ is the line element related to a constant curvature 
"horizon" (D-2)-dimensional manifold.
The horizons are positive simple  roots of the lapse function,
 i.e.
\beq
A(r_H)=0\,,\,\,\,\,\, A'(r_H)\neq 0 \,.
\eeq
The associated Hawking temperature is
\beq
\beta_H=\frac{4\pi}{A'(r_H)}\,.
\eeq
In general, when the extremal solution exists, there exists a critical radius
$r_c$ and we have
\beq
A(r_{c})=0\,,\,\,\,\,\, A'(r_{c})= 0\,,\,\,\,\,\, A''(r_{c})\neq 0 \,.
\eeq
Furthermore, there exists also a relationship between the parameters,
\beq
F(m,g_i)=0\,.
\eeq
When this condition is satisfied, it may happen that the original coordinates
become inappropriate and the metric no more static. Typically, this happens 
in the presence of multiple horizons, namely $\Lambda >0$ and  when 
$A(r)$ has a local maximum, thus
 $A''(r_{c})<0$. 

In order to investigate the extremal limit, it is convenient to introduce
the non-extremal parameter $\epsilon$ and perform the following coordinate 
change 
\beq
r=r_{c}+\epsilon r_1\,,\,\,\,\,\, t=\frac{t_1}{\epsilon}\,. 
\label{gp}
\eeq
and parametrize the non-extremal limit by means of
\beq
F(M,g_i)=k\epsilon^2\,,
\eeq
where the sign of constant $k$ defines the physical range of the black hole
 parameters, namely the ones for which the horizon radius is non negative.
In the near-extremal limit, we may make an expansion for $\epsilon$ small.
As a consequence
\beq
A(r)=A(r_{c})+A'(r_{c})r_1\epsilon+\frac{1}{2}
A''(r_{c})r_1^2\epsilon^2+O(\epsilon^3)\,.
\eeq
In general, when the parameters are near the extremal solution, one
has 
\beq
A(r_{c})=k_1 \epsilon^2\,,\,\,\,\,\,A'(r_{c})=k_2\epsilon^2\,,
\eeq
where $k_i$ are known constants.

Thus, the metric in the extremal limit becomes
\beq
ds^2=-dt_1^2 (k_1+\frac{1}{2}
A''(r_{c})r_1^2)+\frac{dr_1^2}{(k_1+\frac{1}{2}
A''(r_{c})r_1^2)} +r^2_{c}d\Sigma_{D-2}^2\,.
\label{extrem}
\eeq
For $\Lambda <0$ or vanishing, we may have simple horizons and 
 asymptotically AdS or flat space-times. In these cases, it
turns out that the constant $k < 0$ and $A''(r_{c}) > 0$, a local minimum.
Thus, the extremal space-time is locally $AdS_2 \X \Sigma_{D-2}$, where 
$\Sigma_{D-2}$ is a compact constant curvature (horizon) manifold. 

For $\Lambda >0$, we have
the possibility to deal with multiple horizons and 
asymptotically de Sitter  space-times. In these cases,  
$k > 0$ and $A''(r_{c}) < 0$. As a result, the 
 extremal space-time is locally $dS_2 \X S_{D-2}$. 
Let us show this result. Let us  
consider the D-dimensional S-dS space. Recall
\beq
A(r)=1-  \frac{c_DM}{r^{D-3}}-\frac{ r^2}{L^2}\,,
\label{ad1}
\eeq    
and $r_c=\at \frac{D-3}{D-1} \ct^{1/2}L$.
Thus,
\beq
A''(r_c)=-  \frac{2 (D-1)}{L^2}\,.
\label{ad2}
\eeq 
Furthermore, we have near the extremal case
\beq
F=\at 1-\frac{M}{M_c}\ct=k\varepsilon^2
\eeq
with  $k >0$, since $M < M_c$. We also have
\beq
A(r_c)=  \frac{2k}{D-1} \varepsilon\,.
\eeq
As a consequence, $k_1= \frac{2k}{D-1}$ and the extremal metric reads
\beq\label{namet}
ds^2=-\at  \frac{2k}{D-1}- d\frac{r_1^2}{L^2} \ct dt_1^2+
\frac{dr_1^2}{\at  \frac{2k}{D-1}-d \frac{ r_1^2}{L^2} \ct }+
\at\frac{D-3}{D-1}\ct L^2 dS^2_{D-2}\,,
\eeq
namely the  D-dimensional Nariai space-time with topology $dS_2 \X S_{D-2}$.


\begin{thebibliography}{99}


\bibitem{frolov}
V.~P.~Frolov and I.~D.~Novikov,
``Black hole physics'',
Kluwer Academic Publishers (1998).

\bibitem{nollert}
H. P. Nollert,
 Class. Quant. Grav. {\bf 16}, R159 (1999);

\bibitem{kokkotas}
K.D. Kokkotas and B.G. Schmidt, 
Living Rev. Rel. {\bf 2}, 2 (1999).



\bibitem{hod} 
S.~Hod,
Phys.\ Rev.\ Lett.\  {\bf 81},  4293 (1998).


\bibitem{kunstatter} 
G. Kunstatter,
Phys.\ Rev.\ Lett.\  {\bf 90}, 161301 (2003).


\bibitem{beke} 
J.D.~Bekenstein,
Lett.\ Nuovo Cim.\  {\bf 11}, 467 (1974).


\bibitem{seta} 
M. R. Setare,
``Near Extremal Schwarzschild-de Sitter Black Hole Area Spectrum 
from Quasi-normal Modes'', arXiv:hep-th/0401063;
M. R. Setare, `` Non-Rotating BTZ Black Hole Area Spectrum from Quasi-normal 
Modes'', arXiv:hep-th//0311221.


\bibitem{york:83}
J.~W.~York, Phys.~Rev.~{\bf D28}, 2929 (1983).


\bibitem{dreyer} 
O.~Dreyer,
Phys.\ Rev.\ Lett.\  {\bf 90}, 081301 (2003).
 
\bibitem{immirzi} G. Immirzi, 
 Nucl. Phys. Suppl. {\bf 57},
65 (1997).

\bibitem{mann1}
J.~S.~F.~Chan and R.~B.~Mann,
Phys.~Rev.~{\bf D55}, 7546 (1997).

\bibitem{mann2}
J.~S.~F.~Chan and R.~B.~Mann,
Phys.~Rev.~{\bf D59}, 064025 (1999).

\bibitem{cardlemo2001}
V.~Cardoso and J.~P.~S.~Lemos,
Phys.~Rev.~{\bf D63}, 124015 (2001).

\bibitem{horo:00}
G.~T.~Horowitz and V.~E.~Hubeny, Phys.~Rev.~{\bf D62}, 024027 (2000).

\bibitem{CKL:2003}
V.~Cardoso, R.~Konoplya  and J.~P.~S.~Lemos,
Phys.~Rev.~{\bf D68}, 044024 (2003).

\bibitem{bk03}
E.~Berti and K.~D.~Kokkotas, Phys.~Rev.~{\bf D67}, 064020 (2003).

\bibitem{nol} 
H.-P. Nollert,
 Phys. Rev. D {\bf 47}, 5253 (1993).

\bibitem{and}  
N. Andersson, 
Class. Quant. Grav. {\bf 10}, L61 (1993).

\bibitem{bk} 
E.~Berti, K.~D.~Kokkotas,
Phys.~Rev.~{\bf D68}, 044027 (2003).

\bibitem{bckh}
E.~Berti, V.~Cardoso, K.~D.~Kokkotas and H.~Honozawa,
Phys.~Rev.~{\bf D68}, 124018 (2003).


\bibitem{bcy}
E.~Berti, V.~Cardoso, S.~Yoshida, `` Highly Damped Quasinormal 
Modes of Kerr Black Holes: A Complete Numerical Investigation'', 
arXiv:gr-qc/0401052.

\bibitem{motl} 
L.~Motl,
Adv. Theor. Math. Phys.  {\bf 6}, 1135 (2003).

\bibitem{motl1}
L.~Motl and A.~Neitzke,
Adv. Theor. Math. Phys. {\bf 7}, 307 (2003).

\bibitem{neitzke}
A. Neitzke, 
``Greybody factors at large imaginary frequencies'',
arXiv:hep-th/0304080 (2003).

\bibitem{m1}  A.J.M. Medved, D. Martin and M. Visser,
``Dirty black holes: Quasinormal modes'', arXiv:gr-qc/0310009
(2003).

\bibitem{pad} T. Padmanabhan, ``Quasi normal modes: A simple
derivation of the level spacing of the frequencies'',
arXiv: gr-qc/0310027 (2003); T.R. Choudhury and T. Padmanabhan,
``Quasi normal modes in Schwarzschild-DeSitter space-time:
A simple derivation of the level spacing of the frequencies'',
arXiv:gr-qc/0311064 (2003).


\bibitem{car} 
V. Cardoso, J.P.S.~Lemos and S. Yoshida, 
``Quasinormal modes of Schwarzschild black holes in four
and higher dimensions'', arXiv:gr-qc/0309112 (2003).



\bibitem{m2} A.J.M. Medved, D. Martin and M. Visser,
``Dirty black holes: Quasi-normal modes for ``squeezed horizons'',
arXiv:gr-qc/0310097 (2003).




\bibitem{briet}
P. Briet, J. M. Combes and  P.Duclos,
J. Math. Anal. Appl. {\bf 126} 90 (1987).

\bibitem{gerard}
C. Gerard and J. Sjostrand,
Comm. Math. Phys. {\bf 108} 391 (1987).


\bibitem{barreto}
A. Sa' Barreto  and M.~Zworski,
Mathematical Res. Letters. {\bf 4} 103(1997).


\bibitem{reed}
M.~Reed  and B.~Simon,
``Methods of Modern Mathematical Physics'',
IV: Analysis of operators,
Academic Press (1978).


\bibitem{moss}
I.G. Moss and J.P. Norman,
Class. Quantum Grav. {\bf 19 } 2323 (2002).

\bibitem{abdalla}
E. Abdalla and C. Molina,
``Field propagation in the Schwarzschild-de Sitter black hole'',
arXiv:gr-qc/0309078 (2003); 
C. Molina, D. Giugno and E. Abdalla,
``Field propagation in the Reissner-Nordstrom-de Sitter black hole''
arXiv:gr-qc/0309078 (2003); 
 K.H.C. Castello-Branco and E. Abdalla,
``Analytic determination of the asymptotic quasi-normal spectrum of 
 Schwarzschild-de Sitter black holes'',
arXiv:gr-qc/0309090 (2003).

\bibitem{lemos1}
V. Cardoso and J. P.S. Lemos,
Phys. Rev.   {\bf D 67} 084020 (2003).

\bibitem{sun}
V. Suneeta,
Phys. Rev. {\bf D 68} 024020 (2003).


\bibitem{zhidenko}
A. Zhidenko,
Class. Quantum Grav. {\bf 21 } 273 (2004).



\bibitem{futa}
S. Yoshida and T. Futamase,
`Numerical analysis of quasinormal modes in nearly extremal  
Schwarzschild-de Sitter space-time'',
arXiv:gr-qc/0308077 (2003).


\bibitem{kono0}
R. A. Konoplya and A. Zhidenko,
``High overtones of Schwarzschild-de Sitter quasinormal spectrum'',
arXiv:hep-th/0402080 (2004).



\bibitem{halyo:02}
E.~Halyo, ``On the Cardy-Verlinde Formula and the de Sitter/CFT
Correspondence'', JHEP 0203, 009 (2002).

\bibitem{odin}
S.~Nojiri, S.~D.~Odintsov and S.~Ogushi, Int.~J.~Mod.~Phys.~{\bf A17},
4809 (2002).

\bibitem{davies}
P.~C.~Davies, ``Cosmological Horizons and Entropy'',
Class.~Quant.~Grav.~{\bf 5}, 1349 (1988).


\bibitem{caldarelli}
M. Caldarelli, L. Vanzo and S. Zerbini,
``The extremal limit of D-dimensional black holes'',
{\it Geometrical aspects of quantum Fields}, A.A. Bytsenko, A.E. Goncalves, 
B.M. Pimentel Editors, World Scientific (2001) arXiv:hep-th/0008136 (2000).

\bibitem{lemos2}
V. Cardoso, O.J.C. Dias  and J. P.S. Lemos,
``Nariai, Bertotti-Robinson and anti-Nariai solutions in higher dimensions'',
arXiv:hep-th/0401192 (2004).

\bibitem{giappa}
H. Kodama and A. Ishibashi,
Prog. Theor. Phys. {\bf 110}, 701 (2003);
H. Kodama and A. Ishibashi,
Prog. Theor. Phys. {\bf 110}, 901 (2003).


\bibitem{kono1}
R. A. Konoplya, 
Phys. Rev.{\bf D 68} 124017 (2003).

\bibitem{bcg03}
E.~Berti, M.~Cavagli\`a and L.~Gualtieri, ``Gravitational energy loss
in high energy particle collisions: ultrarelativistic plunge into a
multidimensional black hole'', arXiv:hep-th/0309203. 


\bibitem{birm:02}
D.~Birmingham, I.~Sachs and S.~N.~Solodukhin, Phys.~Rev.~Lett.~{\bf
  88}, 151301 (2002).

\bibitem{vish}
C. V. Vishveshwara,
Nature {\bf 22} 936(1970).


\bibitem{ba}
A. Bachelot and A. Motet-Bachelot,
Ann. Inst. H. Poincare' Phys. Theor.  {\bf 59} 3(1993).


\bibitem{new}
R. G. Newton,
{\it Scattering Theory of Waves and Particles},
McGraw-Hill, New Yorl (1966).


\bibitem{will}
B. F. Schutz and  C. M. Will,
Astrophys. J. Lett {\bf 291} L33 (1985);
S. Iyer and C. M. Will,
Phys. Rev.{\bf D  22} 3621(1987).

\bibitem{zasla}
O. B. Zaslavskii,
Phys. Rev.{\bf D 43} 605(1991).
\bibitem{kono}
R. A. Konoplya,
Phys. Rev.{\bf D 68} 024018 (2003).

\bibitem{press}
W. H. Press,
Astrophys. J. {\bf 170} L105 (1971).

\bibitem{ferrari}
V. Ferrari and B. Mashhoon,
Phys. Rev.{\bf D 30} 295 (1984).




\bibitem{GP} 
P. Ginspar and M. J. Perry, 
 Nucl. Phys. {\bf B 222}, 245 (1983).


\bibitem{BH} R. Bousso and S. W. Hawking,  
Phys. Rev. {\bf D 57} 2436 (1998).

 
\bibitem{Z} O.B. Zaslavsky, 
Phys. Rev. Lett. {\bf 76}, 2211 (1996);
 O.B. Zaslavsky, 
 Phys. Rev. {\bf D56}, 2188 (1997).


\bibitem{MS} R. B. Mann and S. N. Solodukhin, 
 Nucl. Phys {\bf B 523} 293 (1998). 



\bibitem{MMS} J. Maldacena,J. Michelson and A. Strominger, 
JHEP {\bf 9902}11 (1999); M. Spradlin and A. Strominger,  
JHEP {\bf 9911} 021 (1999);


\bibitem{NNN} D.J. Navarro, J. Navarro-Salas and P. Navarro, 
 Nucl. Phys. {\bf B 580}, 311 (2000).


\bibitem{banks} 
T.~Banks, 
``Cosmological Breaking of Supersymmetry?''
arXiv:hep-th/0007146.

\bibitem{witten} 
E.~Witten, 
``Quantum Gravity In De Sitter Space''
arXiv:hep-th/0106109.


\end{thebibliography}
\end{document}